\documentclass[%
 reprint,
 amsmath,amssymb,
 aps,
]{revtex4-1}

\usepackage{graphicx}
\usepackage{dcolumn}
\usepackage{bm}

\begin{document}

\preprint{APS/123-QED}

\title{Artificial Tribotactic Microscopic Walkers}

\author{Joshua P. Steimel}
\author{Juan L. Aragones}
\author{Alfredo Alexander-Katz}%
 \email{aalexand@mit.edu}
\affiliation{%
Department of Materials Science and Engineering, Massachusetts Institute of Technology, Cambridge, MA, 02139, USA.\\
}%

\date{\today}

\begin{abstract}
Friction, the resistive force between two surfaces sliding past each other, is at the core of a wide diversity of locomotion schemes. While such schemes are somewhat understood for homogeneous environments, locomotion based on friction in inhomogeneous environments has not received much attention. Here, we introduce and demonstrate the concept of tribotaxis by utilizing microwalkers that detect gradients in the friction coefficient controlled by the density of biological receptors on the substrate. When actuated stochastically, these microwalkers migrate to regions of higher friction, effectively performing chemotaxis. Simulations and theory based on biased random walks are in excellent agreement with experiments. Our results may have important implications in artificial  and natural locomotion in biological environments because interfaces are a prominent motif in nature.
\end{abstract}

\pacs{microwalkers, chemotaxis, random walk}
\maketitle



Tribotaxis is the process by which an active object, biotic or abiotic, detects differences in the effective local friction coefficient and moves to regions of higher or lower friction according to a given protocol. The local friction coefficient between the object and a surface is dictated by the effective interactions between both. Note, however, that due to the inherent directional nature of friction, the friction coefficient can be anisotropic. A prominent example is the skin of many animals that feels rough when stroked in one direction, yet soft in the other. The origin of this asymmetry is the directionality and ordering of hair or scales sticking out from the surface at a slanted angle. Such materials with asymmetric frictional properties are useful to achieve a preferred direction of motion (tribotaxis), and have been shown to be important in multiple processes, such as modulating the effective friction between animal skin (or scales) and a fluid\cite{Bixler:2012jd,Hazel:1999vp}, regulating the flow of complex fluids\cite{Quere:2008cw}, controlling the motion of cells\cite{LeBerre:2013bh}, or skiing up a mountain!  

\begin{figure*}
\centerline{\includegraphics[clip,scale=0.5,angle=-0]{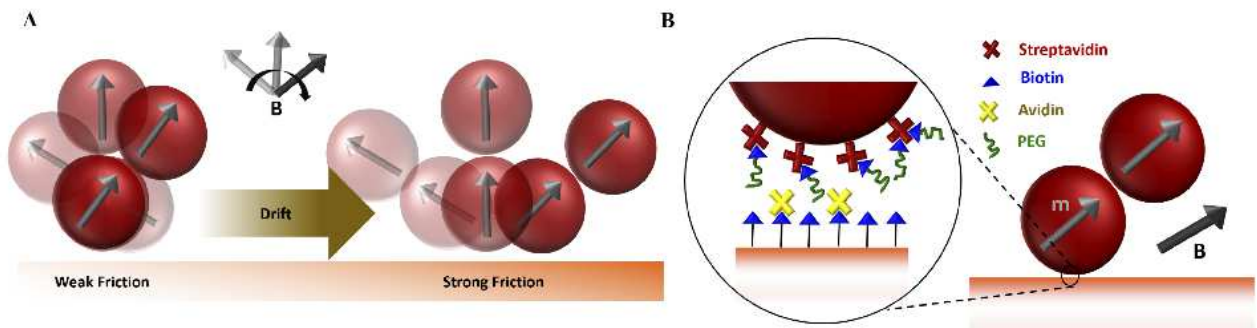}}
\caption{\textbf{Schematic representation of tribotactic microwalkers.} A) A superparamagnetic bead doublet magnetically self-assembled ``walks" on a substrate in the presence of a rotating magnetic field, $\mathbf{B}$. The velocity of walking is determined by the friction with the substrate. When actuated in a random-walk fashion, such tribotactic microwalkers drift to regions of higher friction, which correspond in this work to areas with a higher density of binding sites. B) The effective friction (or interaction) between the beads and the substrate is controlled by functionalizing both surfaces with complementary ligand-receptor pairs. In this work the beads are functionalized with streptavidin and the substrate with biotin.  The beads, 3~$\mu$m in diameter, are further functionalized with mPEG-biotin to mask and reduce the effective binding strength between the walker and the substrate. Free avidin is also used to block available binding sites on the substrate.\label{fig1}}
\end{figure*}

While the motifs that give rise to these asymmetric friction coefficients are rather large, here we envision an alternative microscopic scenario. In particular, one can think of exchanging the mechanical texture by a {\it chemical texture\/} in which friction is dominated by the strength and density of reversible bonds between an object and a substrate\cite{Klein:1996wk,1995Natur.374..607B}. To achieve an effective asymmetric friction coefficient one needs an extended object that can {\it sense\/} differences in friction/binding along its body. The particular system that we have developed is a rotating magnetically assembled colloidal doublet (a microwalker), which converts rotational motion into translational motion when placed near a surface. This is due to the effective friction it experiences with the substrate. Clearly, for weak friction scenarios the doublet primarily slips and translates slowly while for higher friction coefficients the doublet will {\it walk\/} faster (Fig.~\ref{fig1}A). The effective friction for this system can be tuned by functionalizing the surface of the magnetic beads and the substrate with complementary ligand-receptor pairs (Fig.~\ref{fig1}B). Thus, by controlling the spatial density of ligands on the surface it is possible to create gradients in the friction coefficient. When a microwalker is stochastically driven on such gradients its motion resembles that of a biased random walk\cite{Codling:2008bm}. Hence, under such conditions the microwalker drifts toward regions of higher friction, effectively migrating toward areas with a higher ligand density (Fig.~\ref{fig1}A). Hence, this process of friction directed motion can also be understood as a type of chemotaxis-like behavior\cite{Berg:1975wi,Alder:1966wi,Berg:1972wt}. Designing artificial active systems that can perform chemotaxis in an intrinsic way has been a field of intense experimental and theoretical research in the last years\cite{2003JPoSB..41.2755G,Bhattacharya:2009ei,Ebbens:2010fd}. Live cells naturally detect such surface ligand gradients\cite{1967Natur.213..256C}, and this is one of the most important clues for locomotion since cells are constantly encountering surfaces in our bodies. Mimicking such behavior using biological ligand-receptor pairs can potentially allow one to "walk-on" and sense different conditions in the vast amounts of interfaces in tissues and organs. In addition, tribotaxis can be used to locally sense friction in purely synthetic environments.  

\begin{figure}
\centerline{\includegraphics[clip,scale=0.5,angle=-0]{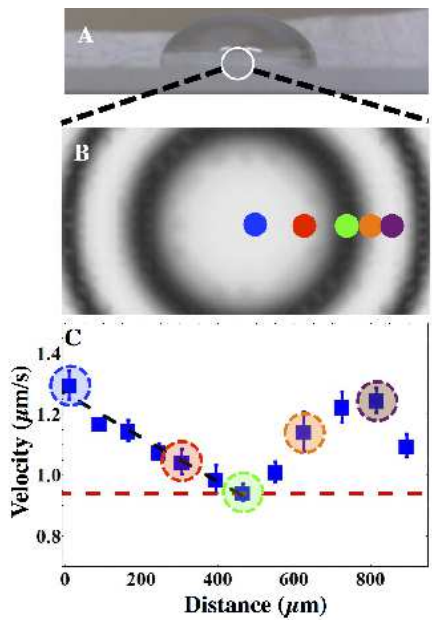}}
\caption{\textbf{Creating gradients in the density of biotin binding sites and mapping the velocity profile of the substrate.} A) A water droplet containing avidin is placed on the biotinylated substrate, and left to evaporate 24hrs. B) Schematic of the gradient in the density of unblocked biotin binding sites. The lighter regions denote a region with a high density of binding sites and darker regions denote regions coated with free avidin ligands. C) Velocity profile ($\mu$m /s) of the microwalkers as a function of the distance to the center of the droplet in $\mu$m. The origin ($x = 0$~$\mu$m) corresponds to the center of the avidin droplet. The velocity of the walkers is mapped across the sample at a rotating frequency of 1~Hz. The horizontal red line denotes the velocity of the walkers on a fully passivated biotin surface, measured in an independent experiment. The colored dots denote the initial positions from which the random walks in Fig.~\ref{fig3} were launched.\label{fig2}}
\end{figure}

Our tribotactic system is composed of two superparamagnetic beads functionalized with streptavidin on their surface. To passivate the surface and reduce the binding strength of the biotin-streptavidin bond between the streptavidin coated beads and the biotin-coated substrate, we decorate such beads with biotinylated PEG, leaving only a few active sites on the surface. The density and distribution of biotin available sites on the substrate can also be screened using free avidin (see Fig.~\ref{fig1}B). To create gradients in friction across the sample, a small water droplet (50~$\mu$l) containing free avidin ligands at a concentration of 1 mg/ml was placed on a hydrophobic biotin substrate and left to evaporate (see Fig.~\ref{fig2}A). The capillarity flows induced by the differential evaporation rates across the droplet carry most of the avidin ligands to the edge, thereby producing a so-called coffee-ring pattern along the perimeter of the droplet\cite{Shmuylovich:2002ej}. Interestingly, an annular deposition pattern of biotin binding sites also develops around the center of the droplet (see Fig.~\ref{fig2}B). Such phenomenon has also been observed in other evaporating systems\cite{Kaya:2010ij}. The resulting frictional patterns exhibit a rugged landscape with areas of high friction separated by low friction "valleys" that are on the order of 1mm. This can be seen in Fig.~\ref{fig2}C, where we plot the velocity of translation of the doublet vs. the distance to the center of the droplet. Note that higher velocities correspond to higher effective friction. In particular, we find the velocity at $x \approx 0$~$\mu$m and $x \approx 800$~$\mu$m to be approximately 30-40 \% larger than at the lowest point, corresponding to $x = 450$~$\mu$m. The velocity displayed in this latter region corresponded to a biotin substrate that was completely coated with avidin ligands (horizontal red dashed line), which was confirmed by an independent measurement on a surface with no free biotin. Assuming a simple Stokes drag scenario, the variations in velocity imply a difference in effective friction forces between both regions of less than $60$~fN.  Thus, our microwalkers are extremely sensitive to minute variations in the friction coefficient. 
For more details about the system, materials and methods we refer the reader to the Supplementary Information (SI) [URL will be inserted by publisher]. 

\begin{figure}
\centerline{\includegraphics[clip,scale=0.4,angle=-0]{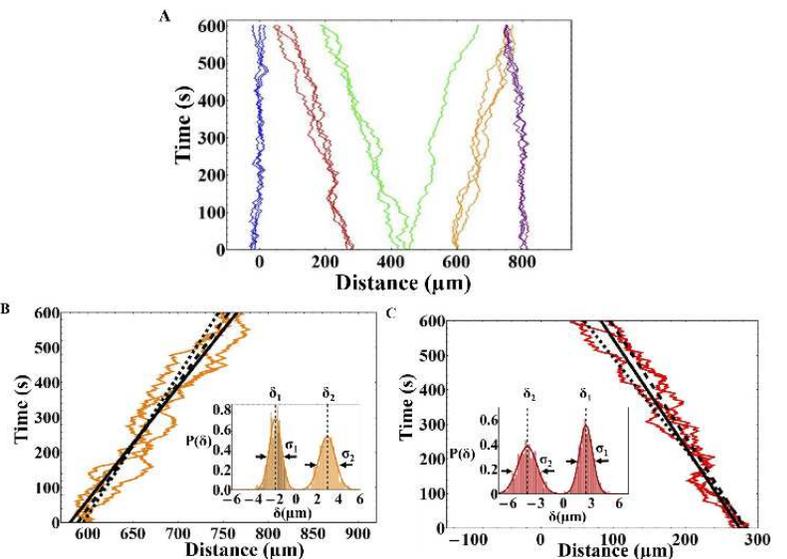}}
\caption{\textbf{Trajectories of tribotactic microwalkers driven stochastically.} A) Random-walk trajectories for doublets initially placed at approximately 0, 300, 450, 600, and 800~$\mu$m from the origin. The color of the walk corresponds to the highlighted position in Fig.~\ref{fig2}C.  A drift towards regions with a higher friction coefficient is observed.  Walkers initially placed in regions with the maximum friction coefficient became trapped. B) Detailed trajectories for walkers initially placed at 600~$\mu$m. The solid, dashed, and dotted lines corresponds to a linear fit to the experimental trajectories (slope 0.28~$\mu$m /s), the simulated trajectories, and the theoretical drift velocity ($u= (\delta_2 - \delta_1) / \Delta t$) respectively. Inset: probability distribution of the displacement in each direction. $\delta_1$ corresponds to the mean displacement in the direction of lower friction, while $\delta_2$ denotes the average displacement toward higher friction areas. C) Trajectories for walkers initially placed at 300~$\mu$m. $\delta_1$ and $\delta_2$ are defined as in part B. The different lines correspond, as in part B, to the linear fits to the experimental (solid line), simulated (dashed line), and theoretical (dotted line) results. The slope of the linear fit to the experimental data is 0.32~$\mu$m / s.\label{fig3}}
\end{figure}

The tribotactic nature of our microwalkers was probed by actuating them with a rotating magnetic field where the direction of rotation is varied in a stochastic fashion, as in a one-dimensional random walk. The total number of steps in each walk is set to 300, and we allow the system to relax for 10 s in between steps, which provides enough time for the doublet to lie parallel to the substrate. The period of actuation, $\tau$, was set at 2 s. In Fig.~\ref{fig3}A, the trajectories for several independent random walks originating from different points in the sample are plotted. As is clear from this plot, the doublets initially positioned at $x=0$ and $x=800$~$\mu$m become {\it trapped\/} in these regions with larger friction coefficient. Walkers initially positioned at regions with a lower coefficient of friction, like $x=300$ and $x=600$~$\mu$m, drift along the gradient in the direction of stronger friction. The region with the lowest friction is at an unstable point and the microwalkers can go either way. This is clearly observed for the walkers initially positioned around $x=450$~$\mu$m, which drifted in both directions (see green trajectories in Fig.~\ref{fig3}A). For microwalkers starting at $x=300$~$\mu$m and $x= 600$~$\mu$m, we computed the drift velocity from the slope of a linear fit to their trajectories (see Fig.~\ref{fig3}B and C). The drift was estimated to be $u =0.32$~$\mu$m/ s ($x=300$~$\mu$m) and $u = 0.28$~$\mu$m /s ($x=600$~$\mu$m). Notice, however, that the drift velocity of these tribotactic microwalkers can be tuned by modifying the parameters of the system, such as the actuation protocol or the size of the walkers. Furthermore, our approach is extremely versatile because the beads can be functionalized with a multitude of biologically relevant motifs, or can even be multiplexed; meaning they can have combinations of motifs to explore cooperative effects or multiple gradients. Thus, such tribotactic microwalkers can be used as sensitive chemotactic probes in biological environments. This overcomes a present challenge that current approaches face in creating artificial chemotactic systems because they typically rely on non-biological fuels\cite{paxton:2004ea,howse:2007ed, Ebbens:2010fd}. 

To further understand our tribotactic system, we analyzed the distances traveled by the walkers at each step for both directions. A probability distribution of such displacements is shown in the insets of Figs. 3B and 3C for the doublets positioned at $x=600$~$\mu$m and $x= 300$~$\mu$m, respectively. From these distributions it is clear that the tribotactic walkers exhibit, on average, asymmetric displacements depending on the direction of motion. The average displacement toward regions with weaker friction is denoted by $\delta_1$ and toward regions with stronger friction is denoted as $\delta_2$, and their corresponding standard deviations are   $\sigma_1$ and $\sigma_2$ (see insets Fig.~\ref{fig3}). It is well known that such asymmetric distributions lead to directed motion (drift). Here we developed a master-equation based model of the stochastic motion exhibited by the doublets that incorporates such distributions of microscopic displacements. By deriving a {\it Fokker-Plank\/} equation via a truncated {\it Kramers-Moyal\/} expansion of the original {\it Master Equation\/}~\cite{VanKampen:2011vs}, it is possible to solve the time evolution of the probability distribution of the position of the walkers. This model equation has been previously used to describe chemotaxis~\cite{Guasto:2012iz,Keller:1971vf}. Note that the actuation protocol is that of a random walk, and thus has no memory. The model and subsequent derivation of the governing equations can be found in the SI [URL will be inserted by publisher]. The final equation that describes the temporal evolution of the probability density is given by,

\begin{equation}
    \centering
\frac{\partial P(x,t)}{\partial t} = - \frac{\partial }{\partial x} [u(x) P(x,t)]+ \frac{1}{2} \frac{\partial^2 }{\partial x^2} [D(x) P(x,t)]
  \label{fokkerplank}
\end{equation}

where $u(x)$ and $D(x)$ are the drift velocity and diffusion coefficient of the walkers, respectively. These coefficients have a microscopic origin, and their functional form can be written in terms of the displacements $\delta_1$ and $\delta_2$ as $u(x)= (\delta_2 - \delta_1) / \tau$ and $D(x) = [(\delta_1^2+ \delta_2^2) + (\sigma_1^2+ \sigma_2^2)] / \tau$. The first term is responsible for tribotactic motion and represents the drift due to the asymmetry in displacements, while the second term corresponds to diffusive motion. The solution to Eq.~(\ref{fokkerplank}) for a constant value of $u$ is a normal distribution whose average drifts with time at a velocity $u$, and the width of the distribution grows as $t^{1/2}$. Such growth can be better seen from simulated random walks with displacements distributed according to those found experimentally (see SI). The drift velocities from such simulated experiments are in excellent agreement with the theoretical and experimental values, as can be seen by comparing the dashed (simulated trajectories), dotted (analytical model), and solid lines (experimental trajectories) in Figs.~\ref{fig3}B and C.   

\begin{figure}
\centerline{\includegraphics[clip,scale=0.4,angle=-0]{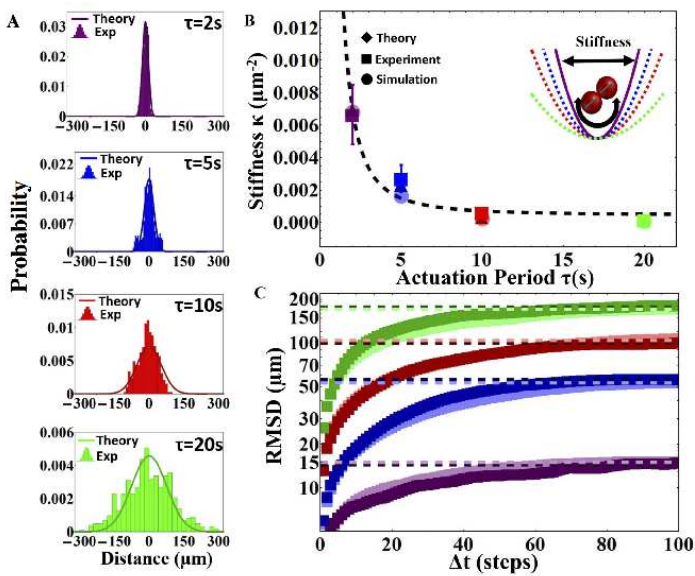}}
\caption{\textbf{Tribotactic trap.} A) Probability distribution of position of tribotactic walkers with an actuation period, $\tau$, of 2s (purple), 5s (blue), 10s (red) and 20s (green), respectively. Solids lines correspond to the numerical solutions to Eq.~(\ref{fokkerplank}) using the experimental parameters for instantaneous motion. B) {\it Stiffness\/} of the tribotactic trap, $\kappa$, computed from the standard deviation of the distributions obtained experimentally (squares), using simulations (circles), and using the analytical model (diamonds). The dashed line represents a fit to the data assuming the functional form $\kappa \propto \tau^{-2}$. C) Root mean squared displacement (RMSD) calculated from the experimental (darker colors) and simulation (lighter colors) trajectories for actuation periods of 2s (purple), 5s (blue), 10s (red) and 20s (green). The RMSDs show the characteristic behavior of a confined random walk.\label{fig4}}
\end{figure}

As described above, the regions of highest friction act as {\it tribotactic traps\/}. To measure the relative strength of these traps with respect to the spatial step size of the random walks, we performed longer stochastic sequences of 500 steps starting at the center of the trap. The effective spatial step size was varied by changing the actuation period, $\tau$, of the rotating magnetic field from 2 s to 5 s, 10 s, and 20 s. The resulting trajectories were analyzed, and the corresponding probability distributions are presented in Figs.~\ref{fig4}A and Extended Data Fig. 5A. The solid lines in Fig.~\ref{fig4}A represent the numerical solutions based on Eq.~(\ref{fokkerplank}) of our tribotactic model, including the directional dependence of the drift term due to the trap. We also simulated many more repetitions of such random walks to get better statistics (see SI). As expected, the increase in $\tau$ translates in to a broadening of the steady-state distributions. In light of the Gaussian-like shape of the steady state distribution, we evaluated the stiffness of the trap $\kappa$ using the standard deviation, $\sigma$, of such distribution from Fig.~\ref{fig4}A. From statistical mechanics we know that the probability distribution of an harmonic oscillator can be written as $P(x) \propto \exp \left( -\frac{1}{2} \kappa(\tau) x^2 \right)$, where $\kappa=1/\sigma^2$. In Fig.~\ref{fig4}B we can see that $\kappa \propto 1/\tau^2$; thus $\tau$ plays the role of a square temperature by analogy to the simple harmonic well. To corroborate the results from the steady-state distributions, we calculated also the root-mean-square displacement (RMSD) for the trajectories. As expected, the RMSD, for both simulations and experiments, exhibit a plateau at long times for all the different $\tau$. Such behavior is characteristic of a confined random walker, and directly corroborates the presence of a tribotactic trap. 

In summary, we have demonstrated tribotaxis using a superparamagnetic doublet functionalized with biological ligands. Such microwalker exploits friction to sense ligand concentration gradients and migrates toward the higher density regions simultaneously,  eventually becoming trapped in such regions. To avoid trapping, we have shown that increasing the actuation time yields a "softening" of the trap and can in principle lead to escape, similar in spirit to the way bacteria escape from their own chemoattractant\cite{Tsori:2007gy}. Thus, the frictional landscapes felt by these tribotactic microwalkers can be tuned, making this system ideal to measure differences in friction coefficients in complex environments or to find areas with relatively higher density of ligands without a priori knowledge. Furthermore, given the similarities between cells and the aforementioned microwalkers in terms of size and the environment in which they move, we believe that nature may already be using friction as an ultra sensitive probe for differences in the density of ligands which ultimately play a determining role in biological processes. 

This work was partially supported by the MITEI BP Fellowship (JPS), the Chang Family (JPS and AAK), and the Department of Energy BES award \# ER46919 (JLA).


\end{document}